\documentclass[mathleft
]{an}
\usepackage{graphicx}

\usepackage{times}
\newcommand{\vsini}{$v$\,sin\,$i$}

\begin{document}

\Pagespan{708}{}
\Yearpublication{2009}%
\Yearsubmission{2009}%
\Month{5}%
\Volume{330}%
\Issue{7}%
 \DOI{10.1002/asna.200911236}%

\title{Magnetic survey of emission line B-type stars with FORS\,1 at the VLT\thanks{Based on observations obtained at the
European Southern Observatory,
Paranal, Chile (ESO programmes 077.D-0406(A), 279.D-5042(A), 380.D-0480(A), 080.D-0383(A)).}}

\author{{S. Hubrig\inst{1}\fnmsep\thanks{Corresponding author: {shubrig@aip.de}}}
\and
M. Sch\"oller\inst{2}
\and
I. Savanov\inst{3}
\and
R.V. Yudin\inst{4,5} 
\and
M.A. Pogodin\inst{4,5}
\and
St. {\v S}tefl\inst{6}
\and
Th. Rivinius\inst{6}
\and
M. Cur\'e\inst{7}
}
\titlerunning{Magnetic survey of emission line B-type stars}
\authorrunning{S. Hubrig et al.}
\institute{
Astrophysikalisches Institut Potsdam, An der Sternwarte 16, 14482 Potsdam, Germany
\and
European Southern Observatory, Karl-Schwarzschild-Str.\ 2, 85748 Garching, Germany
\and
Institute of Astronomy of the Russian Academy of Sciences, Pyatnitskaya St.\ 48, 119017, Moscow, Russia
\and
Pulkovo Observatory, Saint-Petersburg, 196140, Russia
\and
Isaac Newton Institute of Chile, Saint-Petersburg Branch, Russia
\and
European Southern Observatory, Casilla 19001, Santiago 19, Chile
\and
Departamento de F\'isica y Astronom\'ia, Facultad de Ciencias, Universidad de Valpara\'iso, Chile
}

\received{2009 May 25}
\accepted{2009 Jun 30}
\publonline{2009 Jul 20}


\abstract{%
We report the results of our search for magnetic fields in a sample of 16 field Be stars, the binary 
emission-line B-type star $\upsilon$\,Sgr, and in a sample of 
fourteen members of the open young cluster NGC\,3766 in the Carina spiral arm.  The sample of 
cluster members includes
Be stars, normal B-type stars and He-strong/He-weak stars.
Nine Be stars have been studied with magnetic field time series obtained over 
$\sim$1 hour to get an insight into the temporal behaviour and the correlation of magnetic field 
properties with dynamical phenomena taking place in Be star atmospheres.
The spectropolarimetric data were obtained at the European Southern
Observatory with the multi-mode instrument FORS\,1 installed at the 8\,m Kueyen telescope.
We detect weak photospheric magnetic fields in 
four field Be stars, HD\,62367, $\mu$\,Cen, $o$\,Aqr, and $\epsilon$\,Tuc.
The strongest longitudinal magnetic field, $\left<B_{\rm z}\right>$\,=\linebreak[3]\,117$\pm$38\,G, was detected 
in the Be star HD\,62367. Among the Be stars studied with time series, one Be star, $\lambda$\,Eri, 
displays cyclic variability of the magnetic field with a period of 21.12\,min.
The binary star $\upsilon$\,Sgr, in the initial rapid phase of mass exchange between the two 
components with strong emission lines in the visible spectrum,
is a magnetic variable star, 
probably on a timescale of a few months. The maximum longitudinal 
magnetic field $\left<B_{\rm z}\right>$\,=\,$-$102$\pm$10\,G at MJD\,54333.018 was measured using hydrogen lines.
The cluster NGC\,3766 seems to be extremely interesting, where we find evidence for the
presence of a magnetic field in seven early B-type stars out of the observed fourteen cluster members.
}

\maketitle

\section{Introduction}
\label{sect:intro}
The majority of emission line B-type stars constitute so-called Be stars, which are defined as rapidly 
rotating main sequence stars showing normal B-type spectra 
with superposed Balmer line emission.
In addition, these stars are characterized by episodic dissipation and formation of a new
circumstellar (CS) disk-like environment, non-radial pulsations, and
photometric and spectroscopic variability.
A number of physical processes in classical Be stars (e.g., angular momentum transfer to a 
CS disk, channeling stellar wind matter, accumulation of material 
in an equatorial disk, etc.)
are more easily explainable if magnetic fields are invoked (e.g.\ Brown et al.\ \cite{Brown2004}).
Cassinelli et al.\ (\cite{Cassinelli2002}) suggested a Magnetically Torqued Disk model, in which a sufficiently 
strong magnetic field (of the order of 300\,G) channels a flow of wind material towards the 
equatorial plane to form a disk.
Maheswaran (\cite{Maheswaran2003}, \cite{Maheswaran2005}) developed the Magnetic Rotator Wind Disk model,
in which Keplerian disks may be formed by magnetic fields of the order of a few tens of Gauss.
Very recently, Maheswaran \& Cassinelli (\cite{MaheswaranCassinelli2009}) obtained solutions for the structure and evolution 
of a protodisk region, i.e. the disk region that is initially formed when wind material 
is channeled by dipole-type magnetic fields towards the equatorial plane, showing that magnetorotational 
instability may assist in the formation of a quasi-steady disk. According to their calculations, magnetic fields 
of the order of a few tens of Gauss will be able to channel wind flow into a protodisk region.

Due to the high rotation of Be stars and the presence of strong Balmer emission lines, magnetic field measurements
are difficult and rare. The only reported magnetic field detection using 
high-resolution spectropolarimetry  of $\omega$\,Ori (80$\pm$40\,G; Neiner et al.\ \cite{Neiner2003}) was 
not confirmed by recent observations with ESPaDOnS (Grunhut et al., in preparation). 
A longitudinal magnetic field at a level larger than 3$\sigma$ has 
previously been diagnosed in low-resolution spectropolarimetric FORS\,1 observations of the
three Be stars HD\,56014 (=\,EW\,CMa; Hubrig et al.\ \cite{Hubrig2007}), HD\,148184 (=\,$\chi$\,Oph; Hubrig et al.\ \cite{Hubrig2007}),
and in HD\,208057 (=\,16\,Peg; Hubrig et al.\ \cite{Hubrig2006a}).
HD\,208057 has \vsini{}\,=\,104\,km\,s$^{-1}$ and was classified as a Be star by Merrill \& Burwell (\cite{MerrillBurwell1943})
due to the detection of double emission in H$\alpha$.
Recently, Henrichs et al.\ (\cite{Henrichs2009}) confirmed the presence of a magnetic field with new measurements using the 
spectropolarimeters Narval at the T\'elescope Bernard Lyot, France, and ESPaDOnS at the Canada France Hawaii Telescope, during 2007.
However, the presence of emission in  H$\alpha$ was not detected by 
these observations. Thus, the question whether this star is a classical Be star remains open. 

Additional 
spectropolarimetric measurements are needed to firmly establish the presence of weak magnetic fields in Be
stars.
Due to the sparseness of the available magnetic field measurements, it is currently not possible 
to test models, which describe the role of weak magnetic fields in 
launching and stabilizing circumstellar disks in Be stars. 
To obtain constraints on the origin of magnetic fields in early B-type stars,
it is especially important to study the incidence of magnetic fields in members of clusters of different ages.
Here we report the results of our four observing runs with the multi-mode 
instrument FORS\,1 installed at the 8\,m Kueyen telescope at the VLT carried out in the last few years 
with the goal to prove the presence of magnetic fields in a sample of 16 Be stars, 
the binary emission-line B-type star star $\upsilon$\,Sgr,
and fourteen members of the young open cluster NGC\,3766 in the Carina spiral arm. The investigation 
of this cluster, which has a high 
content of Be stars, normal early B-type stars, and He peculiar stars, allows us to draw some first 
conclusions of incidence of magnetic fields in different groups of early B-type stars.

\section{Observations and data reduction}
\label{sect:observations}

\begin{table}[t]
\caption{
Emission line field B-type stars discussed in this paper.
}
\label{tab:targetlist}
\begin{tabular}{rlcl}
\hline\noalign{\smallskip}
\multicolumn{1}{l}{HD} &
\multicolumn{1}{l}{Other} &
\multicolumn{1}{c}{$V$} &
\multicolumn{1}{l}{Spectral} \\
\multicolumn{1}{l}{Number} &
\multicolumn{1}{l}{Name} &
 &
\multicolumn{1}{l}{Type} \\[1.5pt]
\hline\noalign{\smallskip}
33328 & $\lambda$\,Eri &4.2  & B2IVne \\
41335 & V696\,Mon& 5.3 & B2Vne \\
56014 & 27\,CMa  &4.7 & B3IIIe   \\
58011 & NN\,CMa  &7.2 & B1/B2Ib/IInne  \\
58715 & $\beta$\,CMi  &2.9 & B8Ve   \\
58978 &FY\,CMa &5.6 & B1II  \\
60855 & V378\,Pup &5.7 & B2/B3V   \\
62367 & BD$-$04 2062 &7.1 & B9  \\
88661 &QY\,Car &5.8 & B2IVnpe     \\
91465 & p\,Car 	& 3.4 & B4Vne  \\
105435 & $\delta$\,Cen & 2.6& B2IVne \\
120324 & $\mu$\,Cen &3.5 & B2Vnpe\\
127793 &$\eta$\,Cen  &2.3& B1.5Vne \\
148184 & $\chi$\,Oph &4.4& B2Vne  \\
158427 &$\alpha$\,Ara	&2.8 & B2Vne \\
181615 & $\upsilon$\,Sgr &4.6 & B8pI \\
209409 & $o$\,Aqr  & 4.7& B7IVe     \\
224686 & $\epsilon$\,Tuc & 4.6 & B9IV \\[1.5pt]
\hline
\end{tabular}
\end{table}

The observations reported here have been carried out
from 2006 to 2008 in service and visitor mode at the VLT with the FOcal Reducer low dispersion Spectrograph (FORS\,1).
FORS\,1 is a multi-mode instrument equipped with
polarisation analyzing optics comprising super-achromatic half-wave and quarter-wave 
phase retarder plates, and a Wollaston prism with a beam divergence of 22$\arcsec$  in 
standard resolution mode. 
The HD numbers, visual magnitudes and spectral types of the studied emission line 
field B-type stars based on the SIMBAD 
database are listed in Table~\ref{tab:targetlist}. 

Time-resolved series of ten Be stars have been observed in 2006 in service mode
with the GRISM 600B at the resolution $R$\,=\,2000 in the wavelength range 3480--5890\,\AA{}
to cover all hydrogen Balmer lines from H$\beta$ to the Balmer jump. 
We used the Tektronix chip and a non-standard readout mode, A,1$\times$1,low, allowing us
to obtain a signal-to-noise ratio of a few hundred for measurements of the
circular polarisation. Further, the readout time was reduced to about 40\,s by windowing the CCD.

Since we do not know the magnetic field topology on the surface of Be stars,
it is reasonable to study the polarisation induced in the spectral lines
by the Zeeman effect 
across the stellar surface through longer 
time series of exposures with short integration time and with low detection limit.
It is well known that many Be stars exhibit
X-ray flares and rapid variability of absorption features (dimples). 
These features are believed to be produced
from an ablation of photospheric material caused by a nearby flare (e.g.\ Smith et al.\ \cite{Smith1993}).
Hence, it is quite possible that measuring the magnetic field on the stellar surface 
every few minutes, we will be able to study a local transient magnetic field.
In this respect we  note that the
Magnetic Rotator Wind-Disk model 
(Maheswaran \cite{Maheswaran2003})  does not require large-scale organized magnetic fields, axially symmetric fields, or
uniformly strong fields across the entire 
stellar surface. 
This model also qualitatively applies to stars with magnetic fields consisting of flux loops that 
emerge from lower latitudes and thread the disk around the Be star.

For each star we performed time-resolved magnetic field
measurements over one hour (corresponding to a time series of 20 to 30 measurements per star).
In this way we get 
information on the behaviour of the localized transient magnetic field over at least a part of 
the stellar surface. A similar study was conducted for the Be star $\lambda$\,Eri in the past
by Mathys \& Smith (\cite{MathysSmith2000}) with CASPEC at the ESO 3.6\,m telescope,
where some constraints on a possible presence of a magnetic field were discussed.  

The mean longitudinal magnetic field $\left<B_{\rm z}\right>$ is the average over the stellar hemisphere
visible at the time of observation of the component of the magnetic field
parallel to the line of sight, weighted by the local emergent spectral line
intensity.
The detection of a weak magnetic 
field $\left<B_{\rm z}\right>$\,=\linebreak[3]\,136$\pm$16\,G in one of the stars observed with time-series, $\chi$\,Oph, was
already reported in our 
previous work on Be stars (Hubrig et al.\ \cite{Hubrig2007}).
After re-inspection of the previous measurements of $\chi$\,Oph we noted that the detection of 
this field involved a Zeeman feature in the H$\beta$ line filled with strong emission. 
Excluding this line from the measurements we obtain a much weaker magnetic field 
$\left<B_{\rm z}\right>$\,=\linebreak[3]\,17$\pm$18\,G. At the low FORS\,1 resolution of $R$\,=\,2000 it is difficult to decide
whether the observed Zeeman feature is indeed related to the presence of 
a magnetic field or not and its study at a higher resolution 
will be worthwhile. This star is not further considered in this work.

Our measurements revealed that one Be star, $\lambda$\,Eri (=\,HD\,33328), displays 
a cyclic variability of the magnetic field.
While a few other Be stars in our sample appeared variable too, only for $\lambda$\,Eri did the amplitude 
peak stand out nicely. 
Understandably, a confirmation of such a behaviour would 
immediately stimulate a deeper theoretical investigation, and test and further constrain
the recently developed magnetic field models for Be stars. For this reason, we carried out  follow-up observations 
of  $\lambda$\,Eri in 2007 November during two halves of consecutive nights.
For these observations we used GRISM 1200B and an 0\farcs4 slit  ($R$\,=\,4000) to observe the
spectral range 3730--4970\,\AA{}, which includes all Balmer lines 
from H$\beta$ to H12. This time, however, 
no cyclic variability of the magnetic field in $\lambda$\,Eri was detected.
We discuss the possible explanation for the absence of cyclic variability in the Sect.~\ref{sect:timeres}.
Three more Be stars, 27\,CMa, $o$\,Aqr, and $\epsilon$\,Tuc were observed  during this run of two half nights.

\begin{table*}
\caption{
Longitudinal magnetic fields of emission line B-type stars measured with FORS\,1.
All quoted errors are 1$\sigma$ uncertainties.
}
\label{tab:fields}
\begin{tabular}{lccr @{$\pm$} lr @{$\pm$} lc}
\hline\noalign{\smallskip}
\multicolumn{1}{l}{Name} &
\multicolumn{1}{c}{Time} &
\multicolumn{1}{c}{MJD} &
\multicolumn{2}{c}{$\left< B_z\right>_{\rm all}$} &
\multicolumn{2}{c}{$\left< B_z\right>_{\rm hydr}$} &
\multicolumn{1}{c}{Comment} \\[1.5pt]
 &
\multicolumn{1}{c}{Series} &
 &
\multicolumn{2}{c}{[G]} &
\multicolumn{2}{c}{[G]} &
 \\[1.5pt]
\hline\noalign{\smallskip}
$\lambda$\,Eri   & Y & 53955.400       & $-$39        & 33       & $-$13        & 42       & \\
            & Y & 54432.116       & 22           & 11       & 11           & 15       & \\
            & Y & 54433.114       & $-$4         & 11       & $-$6         & 12       & \\
V696\,Mon   &   & 54549.982       & 30           & 40       & 44           & 51       & \\
27\,CMa   &   & 54432.163       & 21           & 20       & 32           & 25       & \\
            &   & 54433.134       & $-$28        & 14       & $-$38        & 17       & \\
            &   & 54549.996       & 60           & 38       & 49           & 43       & \\
NN\,CMa   &   & 54549.082       & 17           & 43       & 30           & 55       & \\
$\beta$\,CMi   &   & 54549.060       & $-$73        & 34       & $-$90        & 41      & \\
FY\,CMa   &   & 54548.995       & 121          & 58       & 70           & 70       & \\
V378\,Pup   &   & 54549.048       & 106          & 46       & 62           & 52       & \\
HD\,62367   &   & 54549.095       & {\bf 99}    & {\bf 32}  & {\bf117}     & {\bf38}  & ND\\
QY\,Car   & Y & 53889.996       & 65           & 53       & 57           & 88       & \\
p\,Car   & Y & 53890.332       & 73           & 96       & 108           & 102      & \\
$\delta$\,Cen  & Y & 53869.232       & 48           & 45       & 85           & 66       & \\
$\mu$\,Cen  & Y & 53869.295       & {\bf $-$80}  & {\bf 24} & 98           & 33       & ND \\
$\eta$\,Cen  & Y & 53862.329       & $-$14        & 28       & $-$5         & 32       & \\
$\alpha$\,Ara  & Y & 53869.353       & $-$5           & 39       & 10           & 45       & \\
$\upsilon$\,Sgr  &   & 54333.018       &{\bf $-$78}   &{\bf 8}   &{\bf $-$102}  & {\bf 10} & CD \\
            &   & 54343.098       & {\bf$-$73}   & {\bf 9}  & {\bf $-$98}  & {\bf 10} & CD\\
            &   & 54361.071       & 19           & 13       & $-$3         & 11       & \\
$o$\,Aqr  & Y & 53955.185       & {\bf 85}     & {\bf 28} & {\bf 98}     & {\bf 31} &ND \\
            &   & 54432.026       & $-$48        & 22       & $-$46        & 31       & \\
            &   & 54433.007       & $-$28        & 22       &$-$13         & 28       & \\
$\epsilon$\,Tuc  & Y & 53869.405       & 9            & 24       & $-$1         & 26       & \\
            &   & 54432.026       & {\bf 74}     &{\bf 24}  & 61           & 28       & ND \\
NGC3766-025 &   & 54550.261 &$-$106        & 55       &$-$75         & 58       & \\
NGC3766-045 &   & 54550.066 & {\bf $-$123} & {\bf 40} &{\bf $-$194}  &{\bf 62}  & CD \\
NGC3766-047 &   & 54549.020 & {\bf $-$134} & {\bf 42} & $-$129          & 58       & CD\\
NGC3766-073 &   & 54550.016 & $-$99        & 48       & $-$115       & 67       & \\
NGC3766-083 &   & 54549.117 & $-$79        & 31       & $-$89        & 34       & \\
NGC3766-094 &   & 54550.327 & {\bf 294}    & {\bf 53} & {\bf 310}    & {\bf 65} & CD \\
NGC3766-170 &   & 54550.186 & {\bf 1522}   & {\bf 34} & {\bf 1559}   &{\bf 38}  & CD\\
NGC3766-196 &   & 54549.151 & $-$11          & 29       &  $-$3         & 40       & \\
NGC3766-200 &   & 54550.375 & {\bf 128}     & {\bf 40}  & {\bf 115}    & {\bf 38}& ND \\
NGC3766-041 &   & 54550.261 &$-$55         & 38       &$-$67         & 52       & \\
NGC3766-055 &   & 54550.066 & 2            & 36       &$-$13         & 45       &  \\
NGC3766-111 &   & 54549.020 & {\bf 112}    & {\bf34}  & 89           & 38       & ND \\
NGC3766-161 &   & 54550.016 & 43           & 29       & 62           & 39       & \\
NGC3766-176 &   & 54550.016 & {\bf 89}     & {\bf 28} & {\bf 120}    &{\bf 34}  & ND \\[1.5pt]
\hline
\end{tabular}
\end{table*}

Hubrig et al.\ (\cite{Hubrig2007})
reported on the presence of a weak magnetic field 
$\left<B_{\rm z}\right>$\,=\linebreak[3]\,38$\pm$10\,G in the binary 
emission-line B-type star $\upsilon$\,Sgr (=\,HD\,181615). Additional three observations of this binary system were 
obtained in service mode during 2007 August and September with GRISM 1200B and an 0\farcs4 slit 

The most recent observations presented here have been carried out in a visitor run on 2008 March 23 and 24 
in the framework of the study of 15 B-type members of the open cluster NGC\,3766. 
The results of this study have in part been reported by McSwain (\cite{McSwain2008}).
Unfortunately, the polarimetric 
spectra of the member star NGC3766-031 were strongly contaminated by a close companion and have not been further 
considered in our study.
During this run, apart from the Be star members of this cluster, we also observed the field Be stars 
V696\,Mon, 27\,CMa, NN\,CMa, $\beta$\,CMi, FY\,CMa, V378\,Pup, and HD\,62367.
These observations have been carried out with the GRISM 600B and a slit 
width of 0\farcs4.

All observations obtained in 2007 and 2008 have been carried out with a new mosaic detector
with blue optimised E2V chips, which was implemented in FORS\,1 in 2007.
It has a  pixel size of 15\,$\mu$m (compared to 24\,$\mu$m for the
previous Tektronix chip) and higher efficiency
in the wavelength range below 6000\,\AA{}.
With the new mosaic detector
and the grism 600B we were able  to cover a much larger
spectral range, from 3250 to 6215\,\AA{}, and from 3680 to 5130\,\AA{} using grism 1200B.
To achieve the highest possible signal-to-noise (S/N) ratio -- as is 
required for accurate measurements of stellar magnetic fields --
the non-standard, 200kHz, low, 1$\times$1, readout mode was used, which makes it possible to 
achieve a S/N  ratio of more than 1000 with only one single exposure.
For each star observed in 2007--2008 we usually took three to five continuous series of two exposures
at the position angles of the retarder waveplate $+$45 and $-$45.
More details on the observing technique with FORS\,1 can be 
found elsewhere (e.g., Hubrig \cite{Hubrig2004a}, \cite{Hubrig2004b}, \cite{Hubrig2008}).

The mean longitudinal magnetic field is diagnosed from the slope of
a linear regression,
\begin{equation}
V/I =
-\frac{g_{\rm eff}\,e}{4\pi{}m_{\rm e}\,c^2} \lambda^2 \frac{1}{I} \frac{{\mathrm d}I}{{\mathrm d}\lambda} \left<B_z\right> + V_0/I_0,
\end{equation}
\noindent
where $V$ is the Stokes parameter which measures the circular polarisation,
$I$ is the Stokes parameter observed in unpolarized light,
$g_{\rm eff}$ is the effective Land\'e factor,
$e$ is the electron charge,
$\lambda$  is the wavelength,
$m_e$ the electron mass,
$c$ the speed of light,
${{\rm d}I/{\rm d}\lambda}$ is the derivative of Stokes~$I$,
and $\left<B_{\rm z}\right>$ is the mean longitudinal magnetic field.
$V_0/I_0$ is a constant term taking into account the remaining instrumental polarization.
Our experience from the study of a large sample of magnetic and non-magnetic
Ap and Bp stars (Hubrig et al.\ \cite{Hubrig2006b}) revealed that this regression technique is very robust
and that detections at a significance level larger than $3\sigma$ 
result only for stars possessing magnetic fields.

Longitudinal magnetic fields were measured in two ways: using only the absorption hydrogen Balmer 
lines or using the whole spectrum including all available absorption lines.
Lines showing evidence for emission were not used in the 
determination of the magnetic field strength.
The feasibility of longitudinal magnetic field measurements in massive stars 
using FORS\,1 in spectropolarimetric mode was demonstrated by recent studies of early B-type stars
(e.g., Hubrig et al.\ \cite{Hubrig2006a}; Hubrig et al.\ \cite{Hubrig2008}; 
Hubrig et al.\ \cite{Hubrig2009}).

\section{Results}
\label{sect:results}

The results of our magnetic field measurements are summarised in Table~\ref{tab:fields}.
In the first three columns we give the name of the targets, indicate whether the star was observed in a longer
time series, and the modified Julian date of the middle of the exposures.
The mean longitudinal magnetic field $\left< B_z\right>_{\rm all}$ measured using all 
absorption lines is presented in Col.~4.
The  mean longitudinal magnetic field  $\left< B_z\right>_{\rm hydr}$ using all hydrogen lines in 
absorption is listed in Col.~5.
All quoted errors are 1$\sigma$ uncertainties.
In Col.~6 we identify new detections by ND and confirmed detections by CD.
We note that all claimed detections have a significance of at
least 3$\sigma$, determined from the formal uncertainties we derive. 
These measurements are indicated in bold face.
While data for stars with extended time series were also analysed using only the pairwise
settings of the retarder waveplate angle of +45$^\circ$ and $-$45$^\circ$ (see Sect.~\ref{sect:timeres}), 
the values of the measured magnetic field presented here were determined using all of the up to 30 exposures.

We detect weak photospheric magnetic fields in 
four field Be stars, HD\,62367, $\mu$\,Cen, $o$\,Aqr, and $\epsilon$\,Tuc.
The largest longitudinal magnetic field, $\left<B_{\rm z}\right>$\,=\linebreak[3]\,117$\pm$38\,G, was detected 
using hydrogen lines in the Be star HD\,62367.
Among the Be stars studied with time series, the Be star $\lambda$\,Eri 
displays cyclic variability of the magnetic field with a period
of 21.12\,min.

The binary star $\upsilon$\,Sgr, in the initial rapid phase of mass exchange between the two 
components with strong emission lines in the visible spectrum
(Koubsk{\'y} et al.\ \cite{Koubsky2006}), is a magnetic variable star, 
probably on a timescale of a few months. The maximum longitudinal 
magnetic field $\left<B_{\rm z}\right>$\,=\linebreak[3]\,$-$110$\pm$10\,G at MJD\,54333.018 was measured using hydrogen lines.

The cluster NGC\,3766 seems to be extremely interesting, where we find evidence for the
presence of a magnetic field in seven early B-type cluster members out of the fourteen members observed.
The strongest magnetic field $\left<B_{\rm z}\right>$\,=\linebreak[3]\,1559$\pm$38\,G was measured in the He-weak star
NGC3766-170, followed by the second strongest magnetic field $\left<B_{\rm z}\right>$\,=\linebreak[3]\,310$\pm$65\,G
measured in the He-strong star NGC3766-094. Among the cluster member Be stars, the strongest 
magnetic field $\left<B_{\rm z}\right>$\,=\linebreak[3]\,$-$134$\pm$42\,G was measured in NGC3766-47.
Surprisingly, magnetic fields of a similar order were also discovered in the normal early B-type stars NGC3766-111 and 
NGC3766-176.

In the following subsections we describe the time-resolved observations of nine Be stars (Sect.~\ref{sect:timeres}),
discuss the results
of the measurements of other stars with magnetic field detections at 3$\sigma $ level (Sect.~\ref{sect:other}),
and in Sect.~\ref{sect:ngc} we  present the magnetic field measurements of members of the young open cluster NGC\,3766.

\subsection{Time-resolved magnetic field measurements: dis-\\ covery of magnetic field cyclic variability 
in $\lambda$\,Eri}
\label{sect:timeres}

\begin{figure}[t]
\includegraphics[width=0.47\textwidth]{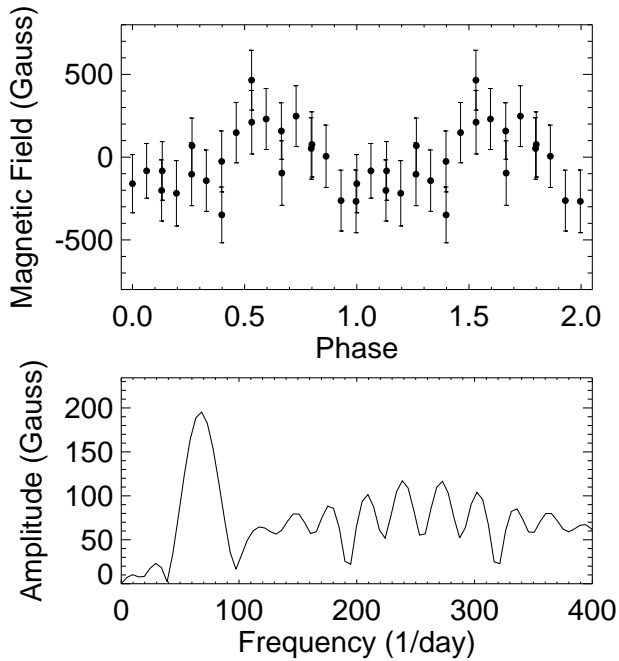}
\caption{
Phase diagram and amplitude spectrum for the magnetic field measurements of 
$\lambda$\,Eri using hydrogen lines in 2006 August.
}
\label{fig:lameri-hydro}
\end{figure}

\begin{figure}[t]
\includegraphics[width=0.47\textwidth]{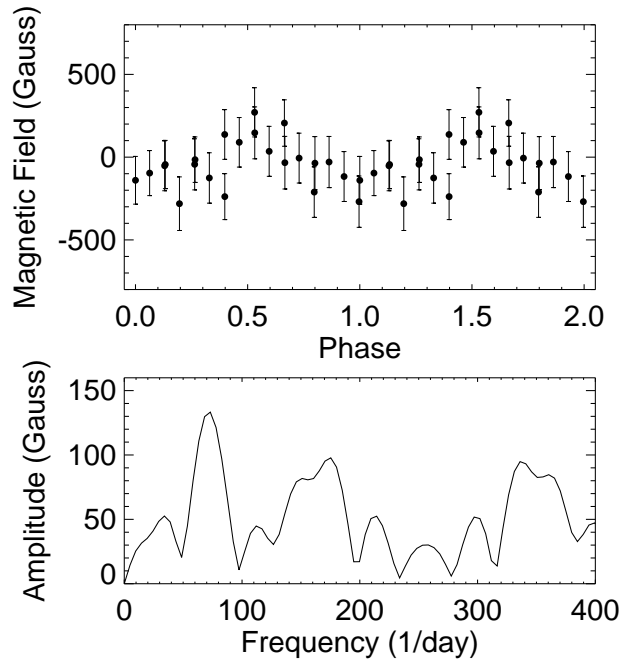}
\caption{
Phase diagram and amplitude spectrum for the magnetic field measurements of 
$\lambda$\,Eri using all lines in 2006 August.
}
\label{fig:lameri-all}
\end{figure}

The nine Be stars with time-resolved magnetic field measurements are very bright objects.
The corresponding integration time for a single measurement of the
magnetic field with the Kueyen 8-m telescope and FORS\,1 accounts for only a few 
seconds. Taking into account time for overheads (retarder waveplate
setting plus readout time), we were able to obtain during one hour 21 consecutive measurements for 
the faintest Be star in our sample (QY\,Car)
and 30 measurements for the brightest Be star in our sample ($\eta$\,Cen). 
The periodicity of the time-resolved magnetic field measurements was analysed by application of 
Breger's code (\cite{Breger1990}).
In the obtained amplitude spectra a 2.4$\sigma$ peak corresponding 
to a period of 21.12\,min was detected in the data set of measurements carried out 
using hydrogen lines in the star $\lambda$\,Eri 
in 2006 August (see lower panel in Fig.~\ref{fig:lameri-hydro}) . 
This peak appears with a 2.2$\sigma$
in the data set of measurements carried out using the whole spectrum (lower panel in Fig.~\ref{fig:lameri-all}). 
The corresponding data sets for measurements using hydrogen lines and the whole spectrum are 
presented in Table~\ref{tab:series_lameri}.
To confirm the detected variability, we repeated our measurements of $\lambda$\,Eri about 16 months later on 
two consecutive nights on 2007 November 27 and 28 over 
few hours on each night to 
sample two different phases of its rotation period. 
However, this time our observations did not reveal any significant periodicity in any of the data sets.
Previous studies of $\lambda$\,Eri indicate that spectral line profiles exhibit short-time periodic variability 
due to non-radial pulsations with a period of 0.7\,days (e.g., Kambe et al.\ \cite{Kambe1993}; 
Rivinius et al.\ \cite{Rivinius2003}). In addition to these line profile variations, 
the He~I $\lambda$6678 line 
is reported to show  dimples with a duration of 2--4\,hours (Smith \cite{Smith1994}). Smith suggests that 
such changes in the line profile are consistent with the stellar rotation rate, as if caused by a rooted 
active spot on the surface. The observed rapid optical line variability develops over tens of minutes or less, 
implying that violent high-energy events occur close to the surface of this star (Smith et al.\ \cite{Smith1997}).
Based on multiwavelength observations in optical, X-rays and FUV, Smith et al.\ (\cite{Smith1997}) proposed that the observed 
violent processes on the surface of this star show great similarity with magnetic flaring. 
Our observations indicate that a strong field could possibly exist locally, but 
with a topology such that its net effect can appear only sporadically in disk-integrated variations.

\begin{figure}[t]
\includegraphics[width=0.47\textwidth]{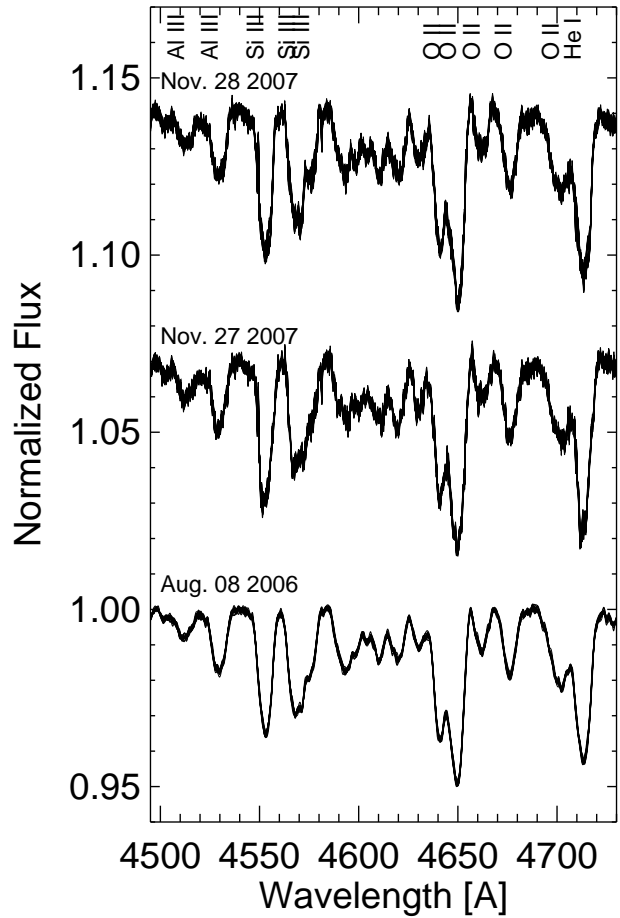}
\caption{
Overlapped spectra of $\lambda$\,Eri for the data sets obtained on 2006 August 8 (\emph{bottom}), 
2007 November 27 (\emph{middle}), and 
2007 November 28 (\emph{top}).
}
\label{fig:lameri_spec}
\end{figure}

\begin{figure}[t]
\includegraphics[width=0.47\textwidth]{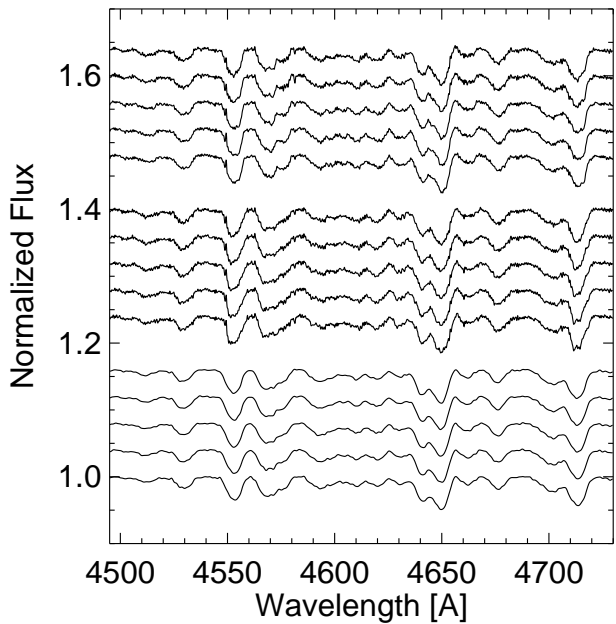}
\caption{
Spectrum variability of $\lambda$\,Eri on three different nights: on 2006 August 8 (\emph{bottom}), 2007 November 27 (\emph{middle}), and 2007
November 28 (\emph{top}).
}
\label{fig:lameri_specvar}
\end{figure}

Trying to understand why we failed to detect any periodicity in our measurements in the follow-up observations, 
we studied the line variations in the polarimetric spectra obtained with FORS\,1 on all 
three observing nights.  Interestingly, while spectral lines in Stokes~$I$ spectra obtained in 2006 August 
appear fairly symmetrical, the spectral lines show rather strong variability in the two time series 
obtained in 2007 November. In Fig.~\ref{fig:lameri_spec} we present for each time series all spectra in the spectral 
region 4500 to 4730 \AA{} overlapped. To show the line profile variations in more detail, we plotted 
in Fig.~\ref{fig:lameri_specvar} five spectra for each time series corresponding to 
equidistant time intervals over the full time used to obtain each data set. The spectra obtained in 2006 August 
are presented in the bottom of the plot, and the spectra obtained on 2007 November 27 and 28 are presented
in the middle and at the top of the plot, respectively.
Asymmetric line profiles are well visible in spectral lines Al\,{\sc iii} $\lambda$4530,
Si\,{\sc iii} $\lambda$4553, O\,{\sc ii} $\lambda$4675, and  He\,{\sc i} $\lambda$4713 in the time series obtained in 2007 November.    
Since the topology of the magnetic field is not known, it is difficult to estimate the impact of non-radial pulsations
causing strong line asymmetries on our measurements. It is quite possible that lines of different elements 
behave differently 
with respect to their pulsation amplitudes and shapes of the line profiles. 
We believe that 
high resolution spectropolarimeters will be more suitable for field measurements in pulsating stars, since 
at higher resolution the Zeeman features in individual lines can be studied separately.  
Our time-resolved magnetic field measurements of the remaining Be
stars indicate that four other Be stars may display a magnetic cyclic variability on the 
time scales of minutes or tens of minutes. The stars QY\,Car,  $\delta$\,Cen, $\alpha$\,Ara, and  $\epsilon$\,Tuc show 
weak signals in the Fourier transforms of our data sets, 
corresponding to periods of 21.86\,min, 27.74\,min, 9.37\,min, and 4.27\,min, respectively.
These stars are good candidates for future time-resolved magnetic field observations 
with high-resolution spectropolarimeters.

\begin{table}
\caption{
Magnetic field time series for $\lambda$\,Eri obtained in 2006 August.
}
\label{tab:series_lameri}
\begin{tabular}{cr @{$\pm$} l r @{$\pm$} l}
\hline\noalign{\smallskip}
\multicolumn{1}{c}{MJD} &
\multicolumn{2}{c}{$\left< B_z\right>_{\rm hydr}$} &
\multicolumn{2}{c}{$\left< B_z\right>_{\rm all}$} \\[1.5pt]
 &
\multicolumn{2}{c}{[G]} &
\multicolumn{2}{c}{[G]} \\[1.5pt]
\hline\noalign{\smallskip}
53955.37939 & $-$160 & 195 & $-$140 & 160 \\
53955.38134 & $-$83  & 197 & $-$45  & 162 \\
53955.38328 & 69     & 186 & $-$15  & 153 \\
53955.38523 & $-$349 & 187 & $-$239 & 154 \\
53955.38718 & 465    & 201 & 271    & 165 \\
53955.38913 & 158    & 190 & 206    & 156 \\
53955.39108 & 52     & 207 & $-$211 & 170 \\
53955.39303 & $-$263 & 204 & $-$117 & 168 \\
53955.39498 & $-$83  & 184 & $-$96  & 151 \\
53955.39693 & $-$218 & 219 & $-$281 & 180 \\
53955.39888 & $-$142 & 206 & $-$126 & 169 \\
53955.40084 & 148    & 203 & 90     & 166 \\
53955.40279 & 230    & 205 & 35     & 168 \\
53955.40475 & 248    & 204 & $-$6   & 167 \\
53955.40671 & 5      & 210 & $-$29  & 172 \\
53955.40867 & $-$267 & 210 & $-$269 & 172 \\
53955.41062 & $-$201 & 205 & $-$52  & 168 \\
53955.41259 & $-$103 & 211 & $-$43  & 172 \\
53955.41455 & $-$26  & 205 & 137    & 167 \\
53955.41652 & 211    & 213 & 147    & 174 \\
53955.41848 & $-$96  & 217 & $-$34  & 177 \\
53955.42045 & 77     & 219 & $-$36  & 178 \\[1.5pt]
\hline
\end{tabular}
\end{table}

\begin{figure}[t]
\vskip-3mm
\includegraphics[width=0.48\textwidth]{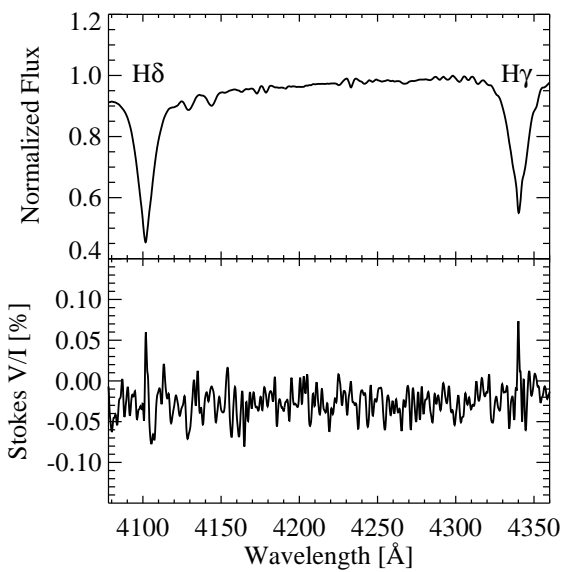}
\caption{
Stokes~$I$ and Stokes~$V$ spectra of $o$\,Aqr in the region including the H$\delta$ and H$\gamma$ lines.
}
\label{fig:Vspec_agr}
\end{figure}

Apart from the study of periodicity in the time series, we used all magnetic field measurements obtained in the time series 
together to determine their average magnetic field.
We obtained field detections at the 3$\sigma$ level for two Be  stars,  $\mu$\,Cen and $o$\,Aqr.
The detected fields are very weak, of the order of one hundred Gauss.
In Fig.~\ref{fig:Vspec_agr} we present Stokes~$I$ and Stokes~$V$ spectra of $o$\,Aqr in the 
spectral region including  the H$\delta$ and H$\gamma$ lines. Noticeable Zeeman features are well visible at the 
positions of both hydrogen lines. Two more observations of this star were obtained during our visitor run 
in 2007 November. The measured magnetic field showed negative polarity at a significance level of only
2.2$\sigma$ on November 27 and of 1.3$\sigma$ on November 28.
 
\subsection{Other emission line B-type stars showing evidence for the presence of a weak magnetic field}
\label{sect:other}

\begin{figure}[t] 
\includegraphics[width=0.48\textwidth]{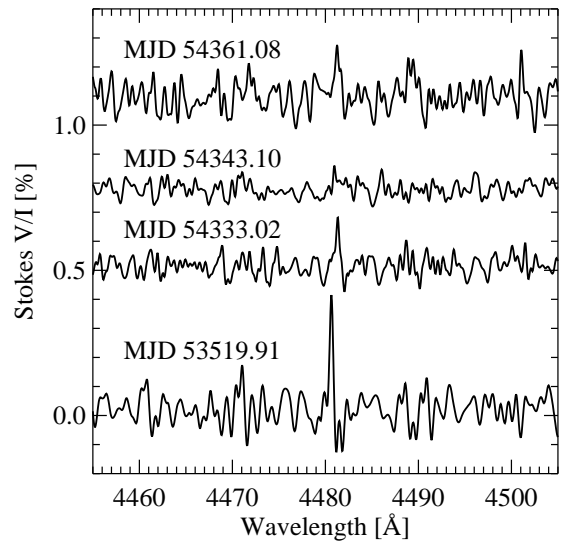} 
\vskip-6mm
\caption{ 
Observed Stokes~$V$ spectra of the emission line star $\upsilon$\,Sgr
over two years in the vicinity of Mg\,{\sc ii} $\lambda$4481. 
} 
\label{fig:Vspec_sgr} 
\end{figure}

\begin{figure}[t]
\includegraphics[width=0.48\textwidth]{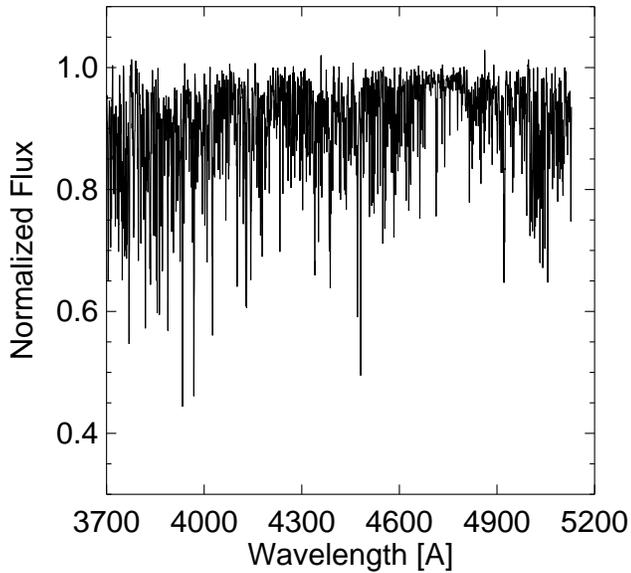}
\caption{
Normalized FORS\,1 Stokes~$I$ spectrum of $\upsilon$\,Sgr.
}
\label{fig:Ispec_sgr}
\end{figure}

\begin{description}
\item[HD\,62367:]
This Be star has only been marginally studied in the past, mainly due to its rather faint magnitude $V$\,=\,7.1. 
It is also one of the 
faintest Be stars in our sample (see Table~\ref{tab:targetlist}). According to Yudin 
(\cite{Yudin2001}) the spectral type of the star is B8e and the $v\sin i$\,=\,114\,km\,s$^{-1}$.
This star was observed only once and exhibits the strongest magnetic field among the
Be stars in our sample.  Using hydrogen Balmer absorption lines we obtained $\left<B_{\rm z}\right>$\,=\linebreak[3]\,117$\pm$38\,G.

\item[$\epsilon$\,Tuc  = HD\,224686:]
This Be star was classified as B8V with $T_{\rm eff}$\,=\linebreak[3]\,13\,000\,K 
and log\,$g$\,=\,3.90
by  Levenhagen \&  Leister (\cite{LevenhagenLeister2006}),
who also determined 
 \vsini{}\,=\,300\ km\,s$^{-1}$.
A weak magnetic field at a 3$\sigma$ significance level, $\left<B_{\rm z}\right>$\,=\linebreak[3]\,74$\pm$24\,G, was detected 
during our observing run in 2007 November.

\item[$\upsilon$\,Sgr = HD\,181615:]
The emission-line star $\upsilon$\,Sgr is a very unusual object, frequently classified as 
a Be star due to the presence of strong emission lines in the visible spectrum. 
It seems to be a magnetic variable star, probably 
on a few months timescale with a maximum longitudinal magnetic field $\left<B_{\rm 
z}\right>$\,=\,$-$102$\pm$10\,G measured in hydrogen lines on MJD\,54333.018. In  Fig.~\ref{fig:Vspec_sgr} we 
present the Stokes~$V$ spectra in the vicinity of Mg\,{\sc ii} $\lambda$4481 taken on four
 different dates over two years.  The evolutionary 
status for this star is not obvious due to the fact that it is a single-line 
spectroscopic binary system currently observed in the initial rapid phase of mass 
exchange between the two components (Koubsk{\'y} et al.\ \cite{Koubsky2006}). 
The star dominating the optical and UV line spectra is less massive and has a spectral 
type B8pI, while the second, almost invisible component is more massive by a factor of 1.57 and has
a spectral type O9V.
The optically visible star is hydrogen poor and the observed spectrum is extremely line rich (see 
Fig.~\ref{fig:Ispec_sgr}).
Hubrig et al.\ (\cite{Hubrig2007}) reported the detection of distinctive Zeeman signatures in the
Ca\,{\sc ii} H\&K lines, which are probably formed in the circumstellar disk around this star. 
Future monitoring of the magnetic field of $\upsilon$\,Sgr over a few 
months with a high resolution spectropolarimeter would be of extreme interest to understand the role 
of the magnetic field in the evolutionary process of mass exchange in a binary 
system. 
\end{description}

\begin{figure}[!ht] 
\vskip-4mm
\includegraphics[width=0.48\textwidth]{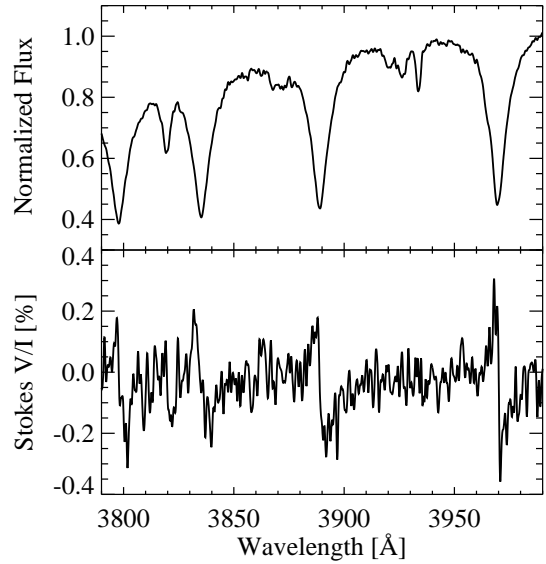} 
\includegraphics[width=0.48\textwidth]{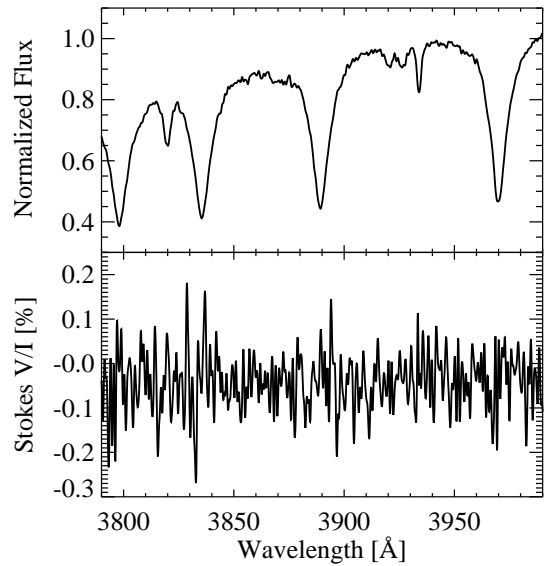} 
\caption{
{\em Upper panel:} Stokes~$I$ and Stokes~$V$ spectra in 
the blue spectral region around high number Balmer lines 
of the He-peculiar member NGC3766-170 of the young open cluster NGC\,3766
with the magnetic field
$\left<B_{\rm z}\right>$\,=\,1559$\pm$38\,G, measured on hydrogen lines. 
{\em Lower panel:} Stokes~$I$ and Stokes~$V$ spectra 
around high number Balmer lines for the candidate Be star NGC3766-45, with a magnetic field
$\left<B_{\rm z}\right>$\,=\,$-$194$\pm$62\,G measured on hydrogen lines.
} 
\label{fig:NGC3766_170} 
\end{figure}

\subsection{Members of the open cluster NGC\,3766}
\label{sect:ngc}

The results of the study of fifteen early B-type
members of the open cluster NGC~3766 have in part been reported by
McSwain (\cite{McSwain2008}), who announced two definite detections in 
He peculiar stars, NGC3766-094 and NGC3766-170, one marginal detection in one Be star, NGC3766-047, and 
one marginal detection in a Be star candidate, NGC3766-045.
A careful treatment of the spectropolarimetric data allowed us to detect weak magnetic fields in 
three additional members of this cluster: in the Be star NGC3766-200 and in the two normal B-type stars NGC3766-111
and NGC3766-176. 
The cluster members with detected magnetic fields are highlighted in
bold face in Table~\ref{tab:fields}.
As was already mentioned in Sect.~\ref{sect:observations}, the polarimetric spectra of NGC3766-031 were contaminated by a close companion
and have not been considered in our study.
As an example, in Fig.~\ref{fig:NGC3766_170} we 
present the observed Stokes~$I$ and Stokes~$V$ profiles of the He peculiar member of this 
cluster and of another cluster member, which was classified as a potential Be star 
by Shobbrook (\cite{Shobbrook1985}) with longitudinal magnetic fields of 
$\left<B_{\rm z}\right>$\,=\linebreak[3]\,+1559$\pm$38\,G and $\left<B_{\rm z}\right>$\,=\linebreak[3]\,$-$194$\pm$62\,G, respectively. 
 
The magnetic fields have been detected in stars with $T_{\rm eff}$ in the range from 
15\,500\,K to 21\,435\,K and log\,$g_{\rm polar}$ from 4.61 to 3.51 (McSwain \cite{McSwain2008}), indicating that 
the presence of a magnetic field is not directly related to the stellar evolutionary phase on the main sequence.

\section{Discussion}
\label{sect:discussion}

Our search for magnetic fields in Be stars revealed that while their
magnetic fields are rather weak, fields of the order of 100\,G and less
are not rare.
Weak magnetic fields are considered to provide a mechanism for launching and stabilizing 
circumstellar disks in Be stars (e.g.\ Brown et al.\ \cite{Brown2008}).
Since a large fraction of stars in our sample was observed only once,
a non-detection of their magnetic field
may be explained by temporal variability of their magnetic fields.
A cyclical variability with a period of 21.12\,min
was detected in one data set of time series in $\lambda$\,Eri, but could not 
be confirmed in the two follow-up time series. 
The cluster NGC~3766 seems to be extremely interesting, where we find clear evidence for the
presence of a magnetic field in seven early B-type cluster members out of fourteen members.

Since magnetic fields can potentially have a strong impact on 
the physics and evolution of B-type stars, it is critical to answer the principal question of the 
possible origin of such magnetic fields. One important step 
would be to conduct observations of members of open clusters and 
associations at different age. To date, we studied the presence of magnetic fields 
only in members of a young open cluster in the Carina spiral arm known for its 
high content of early-B type stars, NGC\,3766, with very surprising results. Along 
with strong magnetic fields detected in He-peculiar stars, weak magnetic fields 
have been detected in a few normal B type stars and in a few Be stars.
We note that the inability to detect magnetic fields in Be stars and normal B-type stars in the past 
is probably related to the weakness of these fields. Future observations will be worthwhile to determine 
the structure of these fields using high signal-to-noise 
spectropolarimetric time series.

\acknowledgements

We are grateful to the referee Dr. J. Madej for useful comments.
MAP and RVY acknowledge the support obtained by RFBR grant No\,07-02-00535a 
and Sci.Schole No\,6110.2008.2, and
MC acknowledges the support obtained by DIUV grant 08/2007.
This research has made use of the SIMBAD database,
operated at CDS, Strasbourg, France.


\label{lastpage}


\begin{thebibliography}{99}

\bibitem[1990]{Breger1990}
Breger, M.: 1990,
Delta Scuti Star Newsletter 2, 21

\bibitem[2004]{Brown2004}
Brown, J.C., Telfer, D., Li, Q., et al.: 2004,
MNRAS 352, 1061

\bibitem[2008]{Brown2008}
Brown, J.C., Cassinelli, J.P., Maheswaran, M.: 2008,
ApJ 688, 1320

\bibitem[2002]{Cassinelli2002}
Cassinelli, J.P., Brown, J.C., Maheswaran, M., et al.: 2002,
ApJ 578, 951

\bibitem[2009]{Henrichs2009}
Henrichs, H.F., Neiner, C., Schnerr, R.S., et al.: 2009,
in: K.G. Strassmeier, A.G. Kosovichev, J. Beckmann (eds.), {\it Cosmic Magnetic Fields: From Planets, to Stars and Galaxies}, IAU Symp. 259, p. 393

\bibitem[2004a]{Hubrig2004a}
Hubrig, S., Kurtz, D.W., Bagnulo, S., et al.: 2004a,
A\&A 415, 661

\bibitem[2004b]{Hubrig2004b}
Hubrig, S., Szeifert, T., Sch\"oller, M., et al.: 2004b,
A\&A 415, 685

\bibitem[2006a]{Hubrig2006a}
Hubrig, S., Briquet, M., Sch{\"o}ller, M., et al.: 2006a,
MNRAS 369, L61

\newpage
\bibitem[2006b]{Hubrig2006b}
Hubrig, S., North, P., Sch{\"o}ller, M., Mathys, G.: 2006b,
AN 327, 289

\bibitem[2007]{Hubrig2007}
Hubrig, S., Yudin, R.V., Pogodin, M., et al.: 2007,
AN 328, 1133

\bibitem[2008]{Hubrig2008}
Hubrig, S., Sch{\"o}ller, M., Schnerr, R.S., et al.: 2008,
A\&A 490, 793

\bibitem[2009]{Hubrig2009}
Hubrig, S., Briquet, M., De Cat, P., et al.: 2009,
AN 330, 317

\bibitem[1993]{Kambe1993}
Kambe, E., Ando, H., Hirata, R., et al.: 1993,
PASP 105, 1222

\bibitem[2006]{Koubsky2006}
Koubsk{\'y}, P., Harmanec, P., Yang, S., et al.: 2006,
A\&A 459, 849

\bibitem[2006]{LevenhagenLeister2006}
Levenhagen, R.S., Leister, N.V.: 2006,
MNRAS 371, 252

\bibitem[2003]{Maheswaran2003}
Maheswaran, M.: 2003,
ApJ 592, 1156

\bibitem[2005]{Maheswaran2005}
Maheswaran, M.: 2005,
in: R.\ Ignace, K.G.\ Gayley (eds.), {\it The Nature and Evolution of Disks Around
Hot Stars},
ASPC\ 337, p. 259

\bibitem[2009]{MaheswaranCassinelli2009}
Maheswaran, M., Cassinelli, J.P.: 2009,
MNRAS 394, 415

\bibitem[2000]{MathysSmith2000}
Mathys, G., Smith, M.A.: 2000,
in: M.A.\ Smith, H.F.\ Henrichs, J.\ Fabregat (eds.),
{\it The Be Phenomenon in Early-Type Stars}, ASPC 214,
p. 316

\bibitem[2008]{McSwain2008}
McSwain, M.V.: 2008,
ApJ 686, 1269

\bibitem[1943]{MerrillBurwell1943}
Merrill, P.W., Burwell, C.G.: 1943,
ApJ 98, 153

\bibitem[2003]{Neiner2003}
Neiner, C., Hubert, A.-M., Fr{\' e}mat, Y., et al.: 2003,
A\&A 409, 275

\bibitem[2003]{Rivinius2003}
Rivinius, T., Baade, D., {\v S}tefl, S.: 2003,
A\&A 411, 229

\bibitem[1985]{Shobbrook1985}
Shobbrook, R.R.: 1985,
MNRAS 212, 591
\bibitem[1994]{Smith1994}
Smith, M.A.: 1994,
in: L.A.\ Balona, H.F.\ Henrichs, J.M.\ Le Contel (eds.),  {\it Pulsation,
Rotation, and Mass Loss in Early-Type Stars},
IAU Symp. 162, p. 241

\bibitem[1993]{Smith1993}
Smith, M.A., Grady, C.A., Peters, G.J., Feigelson, E.D.:  1993,
ApJ 409, L49



\bibitem[1997]{Smith1997}
Smith, M.A., Murakami, T., Ezuka, H., et al.: 1997,
ApJ 481, 479

\bibitem[2001]{Yudin2001}
Yudin, R.V.: 2001,
A\&A 368, 912

\end{thebibliography}
\end{document}